\documentclass[reprint,superscriptaddress, aps,prb,]{revtex4-1}

\usepackage{graphicx}
\usepackage{dcolumn}
\usepackage{bm}
\usepackage{hyperref}

\begin{document}

\title{Weak pinning and vortex bundles in anisotropic Ca$_{10}$(Pt$_4$As$_8$)[(Fe$_{1-x}$Pt$_x$)$_2$As$_2$]$_5$ single crystals}

\author{O. E. Ayala-Valenzuela}
\affiliation{Center for Artificial Low Dimensional Electronic Systems, Institute for Basic Science (IBS), 77 Cheongam-Ro, Pohang 790-784, Korea}

\author{N. Haberkorn}
\affiliation{Centro At\'{o}mico Bariloche, Comisi\'{o}n Nacional de Energ\'{i}a At\'{o}mica. Av., Bariloche, Argentina}
\email[Corresponding author: ]{nhaberk@cab.cnea.gov.ar}

\author{A. B. Karki}
\affiliation{Department of Physics and Astronomy, Louisiana State University, Baton Rouge, Louisiana 70803, USA}

\author{Jisun Kim}
\affiliation{Department of Physics and Astronomy, Louisiana State University, Baton Rouge, Louisiana 70803, USA}

\author{R. Jin}
\affiliation{Department of Physics and Astronomy, Louisiana State University, Baton Rouge, Louisiana 70803, USA}


\author{Jeehoon Kim}
\email[Corresponding author: ]{jeehoon@postech.ac.kr}
\affiliation{Center for Artificial Low Dimensional Electronic Systems, Institute for Basic Science (IBS), 77 Cheongam-Ro, Pohang 790-784, Korea}
\affiliation{Department of Physics, Pohang University of Science and Technology, Pohang 790-784, Korea}

\date{\today}

\begin{abstract}

We report the magnetic field -- temperature ($H-T$) phase diagram of Ca$_{10}$(Pt$_4$As$_8$)[(Fe$_{1-x}$Pt$_x$)$_2$As$_2$]$_5$ ($x\approx 0.05$) single crystals, which consists of normal, vortex liquid, plastic creep and elastic creep phases. The upper critical field anisotropy is determined by a radio frequency technique via the measurements of magnetic penetration depth, $\lambda$. Both, irreversibility line, $H_{irr}(T)$, and flux creep line, $H^{SPM}(T)$, are obtained by measuring the magnetization.  We find that $H_{irr}(T)$ is well described by the Lindemann criterion with parameters similar to those for cuprates, while small $H^{SPM}(T)$ results in a wide plastic creep regime. The flux creep rates in the elastic creep regime are in qualitative agreement with the collective creep theory for random point defects. A gradual crossover from a single vortex to a bundles regime is observed. Moreover, we obtain $\lambda(4~
\text K) = 260(26)$ nm through the direct measurement of the London penetration depth by magnetic force microscopy.

\begin{description}
\item[PACS numbers] {74.25.Dw, 74.25.Ha, 74.25.Wx, 74.25.Op}
\end{description}
\end{abstract}

\maketitle

\section{Introduction}

Since the discovery of high-temperature superconductivity in LaFeAsO$_{1-x}$F$_x$,~\cite{kamihara08}  many new iron-based superconductors (FeSC) have been found and classified into families such as ``122'' (e.g. BaFe$_2$As$_2$),~\cite{rotter08}  ``11'' (e.g. FeSe),~\cite{wang08} and ``111'' (e.g. LiFeAs).~\cite{hsu08} These materials are built from layers of edge-sharing tetrahedra, in which each Fe atom is surrounded by four As or Se atoms. Recently, the family of Ca$_{10}$(Pt$_n$As$_8$)(Fe$_2$As$_2$)$_5$ [with $n = 3$ (10-3-8) and $n = 4$ (10-4-8)] has been reported.~\cite{ni11, kakiya11, lohnert11} They consist of intermediate layers of Pt$_n$As$_8$ stacked in a Ca--Pt$_n$As$_8$--Ca--Fe$_2$As$_2$ sequence. Superconductivity in this system occurs under applied pressure~\cite{gao14} or with chemical doping.~\cite{ni13, thirupathaiah13, cho14} The two different types of intermediate layers ($n = 3$ and $n = 4$) affect significantly the physical properties of these highly anisotropic compounds.~\cite{ni11, shen13} For example, the optimally doped Ca$_{10}$(Pt$_3$As$_8$)[(Fe$_{1-x}$Pt$_x$)$_2$As$_2$]$_5$ has a superconducting transition temperature ($T_c$) around 10 K and an upper critical field of $H_{c2}^c (0) \sim 20$ T.~\cite{watson14} The upper critical field anisotropy ($\gamma=H_{c2}^{ab} / H_{c2}^c $) decreases from 10 close to $T_c$ to 1.5 at low temperatures. For Ca$_{10}$(Pt$_4$As$_8$)[(Fe$_{1-x}$Pt$_x$)$_2$As$_2$]$_5$, $T_c$ can reach 38 K with $x \simeq 0.36$.~\cite{kakiya11} An upper critical field of $H_{c2}^c(0) \sim 92$ T and a change in $\gamma$ from  6-7 close to $T_c$ to $\approx$ 1 at low temperatures has been reported for a sample with $T_c \sim 26$ K and $x \simeq 0.26$.~\cite{mun12} The high $H_{c2}$ and $\gamma$ values in these compounds allow us to study vortex matter with intrinsic high thermal fluctuations.~\cite{ni11, blatter94, kim12} In this context, the  important issue of pinning suppression by thermal fluctuations in materials with random point defects and mixed pinning landscapes stimulates a wide description of the phenomena outside of the cuprate scenario.~\cite{blatter94}

The vortex dynamics of FeSC has been extensively explored.~\cite{prozorov08, tamegai12, beek10} For example, giant flux creep and glassy relaxation have been reported.~\cite{prozorov08} The critical current density ($J_c$) can be enhanced by introducing pinning centers.~\cite{tamegai12} Depending on the applied magnetic field ($H$), $J_c$ is governed by strong defects at low magnetic fields, and by a mixed pinning landscape between strong and weak pinning (random point defects) at intermediate magnetic fields.~\cite{beek10} At high fields a crossover from elastic to plastic creep has been reported in FeSC.~\cite{prozorov08} The $J_c$ values in single crystals of the 10-3-8 and 10-4-8 compounds are smaller than those with similar $T_c$ but smaller anisotropy.~\cite{ding12, tamegai13} This can be attributed to low pinning energies and high thermal fluctuations due to high $\gamma$ values.~\cite{ni13} The irreversibility line or irreversible field ($H_{irr}$) appears in layered superconductors and divides the $H-T$ phase diagram into regions with frozen flux (zero resistance) state and regions with moving flux (finite resistance) in response to an applied current.~\cite{blatter94}

The upper critical field values were obtained in magnetic fields up to 60 T. The $\gamma$ value changes from $\approx 5$ around $T_c$ to $\approx 3$ at 20 K. The irreversibility line can be fit with the Lindemann criteria by considering $c_L = 0.2$ and a Ginzburg number $Gi \approx 0.016$. We find a crossover field at low magnetic fields where the flux creep rate decreases when the magnetic field increases. This can be understood by considering the weak collective pinning scenario (random point defects) where a change in the vortex dynamics is expected in the crossover between the single vortex and vortex bundle regimes. When the magnetic field is increased, the vortex dynamics presents a crossover from elastic to plastic at a field much smaller than the upper critical field and below the irreversibility line. In addition, the absolute value of the magnetic penetration depth $\lambda (4~\text K) = 260(26)$ nm is determined by magnetic force microscopy (MFM).

\section{Experimental details}

High quality Ca$_{10}$(Pt$_4$As$_8$)((Fe$_{1-x}$Pt$_x$)$_2$As$_2$)$_5$ ($x \approx 0.05$) single crystals were grown by the flux method.~\cite{karky} For measuring the penetration depth ($\lambda$), MFM measurements were carried out: two Meissner response curves, one from the 10-4-8 single crystal and one from a Nb reference film, are directly compared in a single cool-down.~\cite{kim-ssc12} The magnetic field dependence of radio frequency (rf) contactless penetration depth, known as proximity detection oscillator (PDO),~\cite{altarawneh09} was measured by applying magnetic fields up to 60 T both parallel ($H//ab$) and perpendicular ($H//c$) to the $ab$ plane of the sample in a pulsed magnet system. The rf technique is a sensitive and accurate method for determining the $H_{c2}$ of superconductors, especially in pulsed magnetic fields.~\cite{mielke01} The details of the $H_{c2}$ determination are described in Ref.[~\cite{mun12}]. Magnetization measurements were performed in a commercial superconducting quantum interference device (SQUID) magnetometer. The critical current density was estimated by applying the Bean critical-state model~\cite{bean64} to the magnetization data, obtained in hysteresis loops, which lead to $J_c=\frac{20 \Delta m}{dw^2 (l-w/3)}$, where $\Delta m$ is the difference in magnetization between the upper and lower branches of the hysteresis loop, and $d=0.2$ mm, $w=1.5$ mm, and $l=2$ mm are the thickness, width, and length of the sample ($l > w$), respectively. The flux creep rate, $S=-\frac{d (\mathrm{ln}⁡ J_c)}{d (\mathrm{ln}⁡ t)} $, was recorded over periods of one hour (with $t$ being the time). The initial time was adjusted considering the best correlation factor in the log-log fitting of the $J_c(t)$ dependence. The initial critical state for each creep measurement was generated by applying a field variation of $H \approx 4 H^*$, where $H^*$ is the field for the full-flux penetration.~\cite{yeshurun96} Deviations from the logarithmic decay in relaxation measured over a long period of time were adjusted by considering the interpolation equation (see description below).~\cite{blatter94}

\section{Experimental Results and Discussion}

We will first focus on the determination of the magnetic penetration depth $\lambda$ and the coherence length $\xi$, before moving on to vortex dynamics. In superconducting single crystals and films whose thickness exceeds $\lambda$, the Meissner response force acting upon a magnetic force microscopy tip obeys a universal power-law dependence $F(z) \sim (z+\lambda)^{-2}$ in the monopole approximation.~\cite{xu95, coffey95} The frequency shift of the tip resonance is proportional to the gradient of the force, i.e., $\delta f \sim dF(z)/dz$. Therefore, by shifting the Meissner data of the 10-4-8 sample with respect to that of Nb along the z axis (in order to overlay one another), one can obtain $\lambda$ of 10-4-8: $\lambda_{10-4-8}(T)= \lambda_{Nb}(T)+ \delta \lambda(T)$, where $\delta \lambda$ is the magnitude of the shift $\delta z$. The difference $\delta \lambda$ between Nb and 10-4-8 is 150 nm, resulting in $\lambda_{10-4-8}$ (4 K) $= 260(30)$ nm. Our experimental error is around 10 \%, resulting from the overlay process of the two Meissner curves of the different $\lambda$ values. The uncertainty of the $\lambda$ extrapolation from 4 K to 0 K is as small as few percent, and thus negligible compared to the main source of the uncertainty.~\cite{cho14}

The 10-4-8 phase Ca$_{10}$(Pt$_{4- \delta}$As$_8$)[(Fe$_{0.97}$Pt$_{0.03}$)$_2$As$_2$]$_5$ with $\delta \approx 0.246$ and  $T_c = 26.5$ K shows $H_{c2}$(0) values of around 90 T for $H//c$ and 92 T for $H//ab$.~\cite{mun12} The upper critical field anisotropy thus changes from $\approx$ 7 close to $T_c$ to $\approx$ 1 at low temperatures. By using these values, we obtain the coherence length $\xi_{ab}(0)=1.90 \pm 0.02$ nm and $\xi_c (0) = 1.85 \pm 0.01$ nm from $H^c_{c2}= \Phi_0/[2 \pi \xi_{ab}^2(0)]$ and $H^{ab}_{c2}=\Phi_0 / [2 \pi \xi_{ab}(0) \xi_c(0)]$ (with $\Phi_0$ the single quantum flux), respectively. As discussed below, the $H_{c2}(T)$ in our sample presents similar dependences to those reported in Ref.[~\cite{mun12}]. Small changes in $\xi_{ab}(0)$ and $\xi_c(0)$ in our sample compared to Ref.[~\cite{mun12}] are expected considering the difference in superconducting transition temperature (i.e. $H_{c2}$ at 15 K is $\approx$ 10 \% higher than that in Ref.[~\cite{mun12}]). By using $\xi_{ab}(0) = 1.9(1)$ nm and $\lambda_{ab}(0) = 260 (26)$ nm, we obtain the thermodynamic critical field $H_c (T=0~\text K)=\frac{ \Phi_0}{2\sqrt{2}\pi \lambda \xi} $=   4700 (400) Oe. These values are related to a theoretical depairing current $J_c (T=0~\text K)=\frac{ cH_c}{3\sqrt{6}\pi \lambda } $=  77 (7) MAcm$^{-2}$ (with $c$ the light speed). The strength of the order parameter thermal fluctuations, estimated by the Ginzburg number $Gi=\frac {1}{2} [ \frac {\gamma T_c }{H_c^2 (0)\xi^3 (0)}]^2 \approx 0.016$, is of the same order of magnitude as that for YBa$_2$Cu$_3$O$_7$ (YBCO).~\cite{blatter94} The Ginzburg number $Gi$ is related to the width of the critical regime, $\Delta T_c > Gi \cdot T_c$, and the melting line ($B_m$) $B_m = \frac{5.6 c_L^4}{Gi} H_{c2}(0)(1-T/T_c)^2$ (here $c_L=0.2$ is the Lindemann number).~\cite{blatter94}

\begin{figure}[tbp]
\label{figure1}
\centering
\includegraphics[width=8cm]{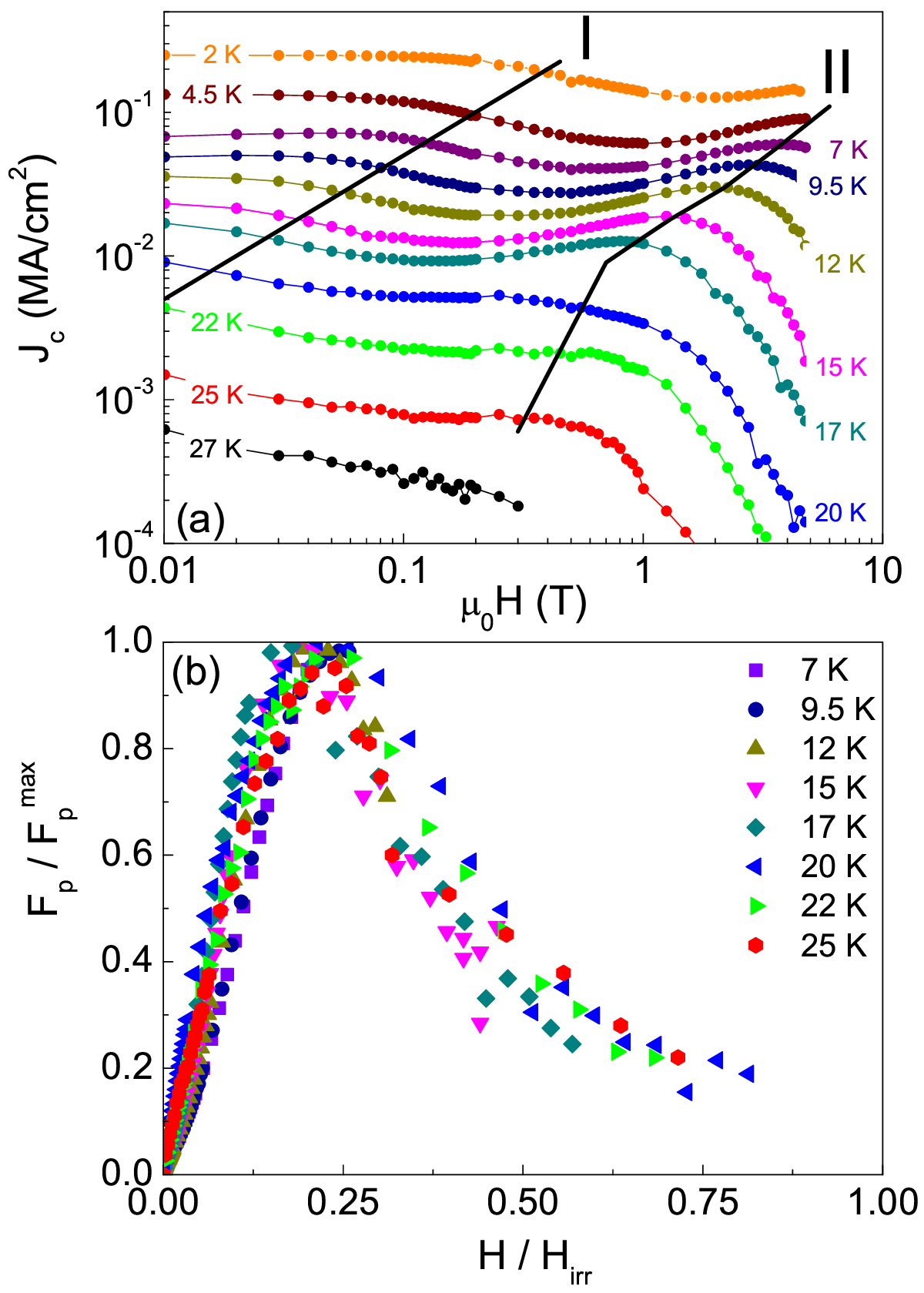}
\caption{(Color online) (a) Magnetic field dependence of the critical current density ($J_c$) for the 10-4-8 single crystal at different temperatures. Lines I and II indicate the regions affected by self-field and the maximum $J_c$ at the fish tail, respectively. (b) Normalized pinning force ($Fp$) versus normalized magnetic field ($h =H/H_{irr}(T)$) at temperatures between 7.5 K and 25 K.}
\end{figure}

In order to study the vortex dynamics, we performed magnetization experiments. Fig. 1(a) shows the critical current density ($J_c$) as a function of the magnetic field ($H$) at different temperatures. Line I marks the self-field regime (SF, $H^* = J_c \cdot d$). Here, the magnetization is affected by geometrical barriers.~\cite{daeumling91} Line II identifies the fishtail or second peak in the magnetization (SPM). This crossover line ($H^{SPM}$) has been extensively discussed and was associated with a change in the vortex dynamics from elastic to plastic creep.~\cite{prozorov08, abulafia96} Note that below 17 K, $J_c(H)$ increases with increasing $H$ between the lines I and II. This can be attributed to a change in the pinning landscape by the magnetic field due to the suppression of superconductivity around point defects.~\cite{das11} In this context, the SPM is strongly reduced by removing inhomogeneities in the samples.~\cite{oka00, haberkorn} However, similar features in the $J_c(H)$ dependence of clean YBCO single crystals have been discussed in the context of a crossover between single vortex and vortex bundle relaxation.~\cite{krusin92} The absence of a fishtail at $T > 17$ K could be related to strong reduction in the pinning (point defects) by thermal smearing (depinning temperature). In highly anisotropic materials, the thermal softening of the vortex core pinning will occur when fluctuations become comparable to $\xi$, and thus the pinning landscape is smeared out on the same length scale.~\cite{watson14} In addition, the anomalous $H_{c2}(T)$ dependence produces a strong change in $\xi$. For example, considering $H^c_{c2}$ (17 K) $\approx 23$ T we obtain $\xi_c$ (17 K) = 3.8(1) nm, which reduces the individual pinning force of a random point defect in the collective pinning in a similar way to that in cuprates due to the increase in $\xi$ and the thermally activated depinning.~\cite{krusin92} Similar features in $J_c(H)$ related to the absence of SPM have been observed in Co doped BaFe$_2$As$_2$ above 20 K.~\cite{prozorov08, kim15} We have not observed a power law behavior, $J_c$ vs. $H^{-\alpha}$, which is usually associated with pinning dominated by strong pinning centers ($\sqrt{2} \xi (T) <$ defect radius).~\cite{beek02} On the other hand, the absolute $J_c(H=0)$ values have strong temperature dependence with $\approx 0.25$ MAcm$^{-2}$ at 2 K and $\approx 0.12$ MAcm$^{-2}$ at 4.5 K. The $J_c(H=0)$ at 2 K is $\approx 0.3$ \% of $J_0$, which is consistent with the weak collective pinning (weak disorder potential with $J_c$$\ll$$J_0$).~\cite{blatter94} The $J_c$ curves and the flux creep rate ($S$) between the SF and the SPM are well described by considering single vortex pinning (see discussion below) and relaxation by vortex bundles.~\cite{civale94} In this context, the increase of $J_c$ values at intermediate fields can be related to strong changes in the vortex dynamics and a reduction in the vortex relaxation.

In order to verify the pinning mechanism at different temperatures we analyze the pinning force ($F_p = J_c H$). In conventional superconductors, $F_p(H, T)$ scales as $F_p/F_{p,max} = Ah^m(1 - h)^l$ , where $F_{p,max}$ is the maximum $F_p(H)$ at each temperature, $A$ is a constant, $m$ and $l$ are exponents that depend on the pinning mechanism, and $h=H/H_{c2}(T)$.~\cite{dew74} This relationship cannot be directly applied here because it is only valid for $h$ defined by $H_{c2}(T)$ rather than $H_{irr}(T)$ (see discussion below). However, our analysis allows a comparison of pinning forces at different temperatures.  Fig. 1(b) shows the $F_p /F_{p,max}$ vs. $h$ (defined as $h=H/H_{irr}$) at different temperatures. The $H_{irr}$ line was determined from irreversible magnetization ($T > 20$ K) and by using the Lindemann criterion ($T < 20$ K, see discussion below). The different curves at $T > 7.5$ K show the maximum pinning force at $h \approx 0.25$. This value is similar to the one obtained in Na doped CaFe$_2$As$_2$ single crystals with random point defects introduced by proton irradiation ($\sqrt{2} \xi (T) >$ defect radius).~\cite{haberkorn14}

\begin{figure}[tbp]
\label{figure2}
\centering
\includegraphics[width=8cm]{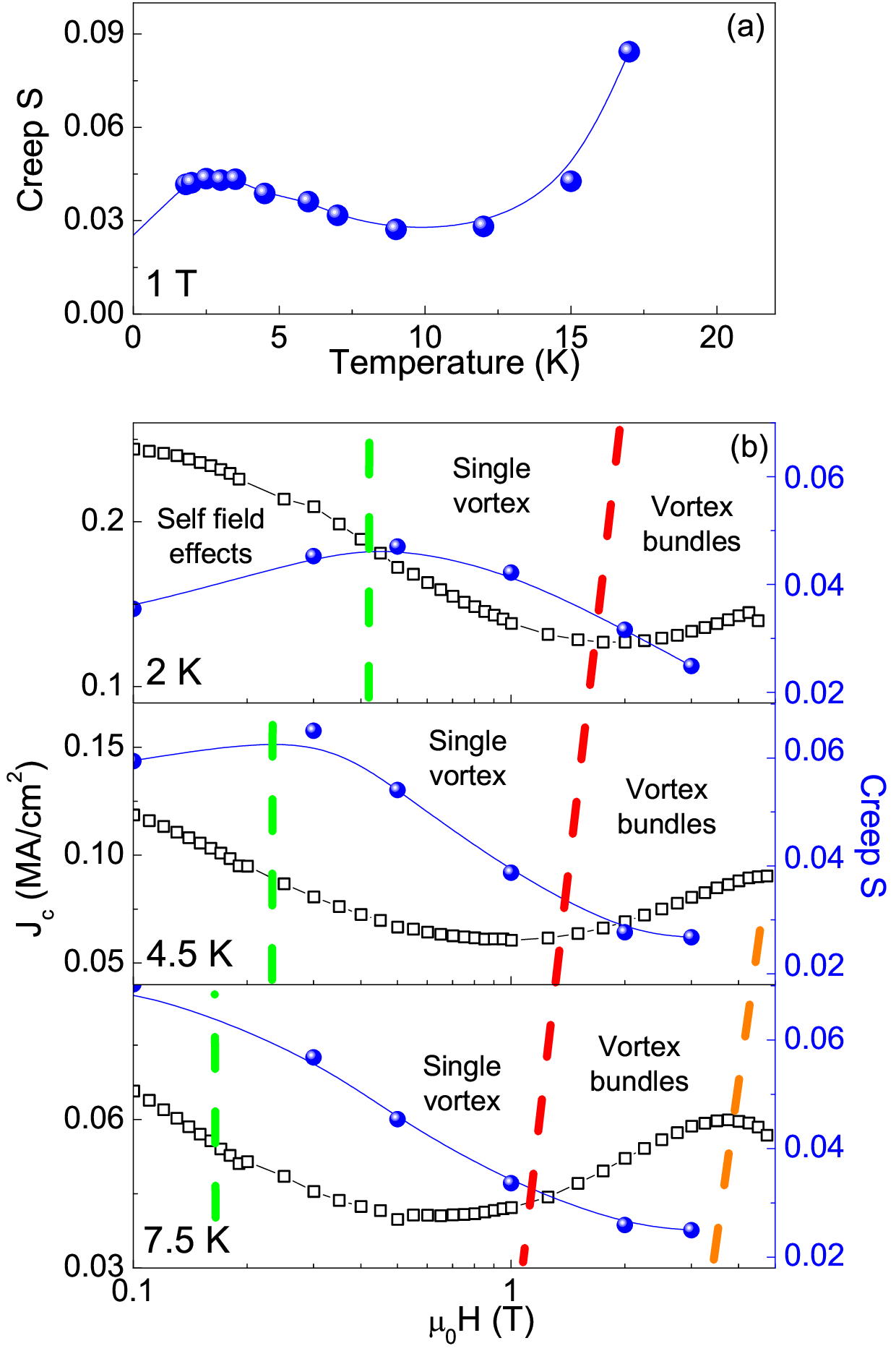}
\caption{(Color online) (a) Temperature dependence of the flux creep rate ($S$) at $\mu_0 H = 1$ T. (b) Magnetic field dependence of $J_c$ (black boxes) and $S$ (blue spheres) at 2 K, 4.5 K, and 7.5 K. The dotted lines are guides for the eye to show the crossover from the single vortex to bundle regime and from vortex bundle to plastic creep.}
\end{figure}

To further understand the collective pinning scenario, we performed creep relaxation measurements at different temperatures and magnetic fields.~\cite{civale94} Fig. 2(a) shows the temperature dependence of the flux relaxation rate $S$ at $\mu_0H = 1$ T. The $S$ values at low temperatures suggest the presence of quantum creep (see fit line Fig. 2 (a)).~\cite{klein14} The theoretical value for the quantum creep can be estimated by $S^Q\cong \frac {e^2}{\hbar} \frac {\rho _n }{\xi} \sqrt \frac {J_c}{J_0}$ (where $\rho_n$ is the resistivity in the normal state). The theoretical prediction $S^Q \approx 0.02$ is in good agreement with the extrapolation to zero temperature from the experiment (see Fig. 2 (a)). For the estimation we use $\rho_n = 0.25$ m$\Omega$cm and $\xi_{ab}(0) = 1.9$ nm.~\cite{mun12, karky} On the other hand, the $S$ value presents a modulation in temperature and a crossover to fast (plastic) creep at $T > 15$ K. The modulation in $S(T)$ can be understood as a modulation in pinning energy or a change in the vortex dynamics by considering single vortex and vortex bundle relaxation. Fig. 2 (b) shows a comparison between the $J_c(H)$ and $S(H)$ at 2 K, 4.5 K, and 7.5 K, respectively. The self-field region was excluded from the analysis.~\cite{haberkorn11} Note that between the SF and the SPM the $S$ values are remarkably large at low $H$, and their minimum at intermediate fields shifts systematically. The collective pinning theory predicts~\cite{blatter94}
\begin{equation}
S=-\frac{T}{U_0 + \mu T \mathrm{ln}⁡ (t/t_0)}
\end{equation}
where $U_0$ is the collective barrier in the absence of a driving force, $\mu$ is the glassy exponent and $t_0$ is an effective hopping attempt time. The nature of the vortex structure and the vortex pinning mechanisms can be inferred from $\mu$, which scales the effective energy barrier and the persistent current density ($J$) as
\begin{equation}
U(J) \approx U_0 \left(\frac{J_c}{J}\right)^{\mu}
\end{equation}
and depends on the creep regime. The glassy exponent $\mu$ depends on the dimension and length scales for the vortex lattice in the collective-pinning model. For random point defects in the three-dimensional case $\mu$ is 1/7 (single vortex, SVR), 3/2 or 5/2 (small vortex bundles, $sb$), and 7/9 (large vortex bundles, $lb$).~\cite{blatter94} The critical current-density ratio ($J/J_0$) is the fundamental quantity characterizing the strength of the disorder potential. The SVR corresponds to weak fields where the distance between the vortex lines is large and their interaction is small compared to the interaction between the vortices and the quenched random potential. The SVR occurs at low fields during the initial stage of the relaxation when $J < J_c$. Single-vortex collective pinning is expected as long as $\gamma L_c <$ inter-vortex distance ($a_0$) ($L_c$ is the Larkin length, $L^c_c = \gamma^{-1} \xi \sqrt{J_0 / J_c}$). Above $\gamma L_c \approx a_0$ the vortices begin to interact as the relaxation slows down. This regime is associated with relaxation by vortex bundles, and a new crossover from $\mu = 3/2 (sb)$ to $\mu = 7/9 (lb)$ is expected. The $U_c (H//c)$ in the SVR can be estimated by $U_c^{SVR} \approx \frac{H_c^2}{\gamma \xi^3} \sqrt{J_c^c/J_0} \approx T_c \sqrt{\frac{J_c^c (1-T/T_c)}{J_0 Gi}}$.~\cite{blatter94} Considering $J_c(T = 0) \approx$ 0.25 MAcm$^{-2}$ and $J_0 = 77$ MAcm$^{-2}$, we obtain $U_c^{SVR}(0) \approx 15$ K.  The crossover between SVR and vortex bundles (in an anisotropic superconductor with $H//c$) is expected at $B_{sb} = \beta_{sb} \frac{J_c}{J_0}H_{c2}$, with $\beta_{sb} \approx 5$.~\cite{blatter94} Based on our experimental results, the SVR should be extended to $\mu_0H \approx 1$ T at 0 K. The $B_{sb}(T)$ dependence is given by $\beta_{sb} \approx H_c \left[1+\frac{T}{\widetilde {T_{dp}^S} }\right]\mathrm{exp}⁡\left[-2c(\alpha+\frac{T}{\widetilde {T_{dp}^S} })^3\right]$, where $c$ and $\alpha (\approx 1)$  are constants and $~{T}_{dp}^S$  is the depinning temperature (SVR is suppressed by thermal fluctuations). On the other hand, a crossover from $sb$ to $lb$ is expected at $B_{lb} (0)\approx \beta_{lb} H_{c2} (\frac{J_{SVR}}{J_0})[\mathrm{ln}⁡ ⁡(\gamma \kappa \frac {J_{SVR}}{J_0})]^{2/3}$, which with $\beta_{lb}= 2$, $\gamma \approx 1-2$ [ref.~\cite{mun12}] and $\kappa= \lambda_{ab}(0)/ \xi_{ab}(0) \approx 130$ corresponds to $B_{lb} \approx 1.5- 1.8$ T. The prediction corresponds to a narrow $sb$ regime still at 0 K.~\cite{blatter94} By using $\mathrm{ln}$($t/t_0$) $\approx$ 29,~\cite{civale94} and considering negligible $U_0$, plateaus [$S=1/\mu \mathrm{ln} (t/t_0)$] with $S \approx 0.022 (sb)$ and $S \approx 0.042 (lb)$ are expected. These values are smaller than the experimental observations above the self-field regions (i. e. $S \approx 0.05$ (2 K, 0.5 T), $S \approx 0.06$ (4.5 K, 0.2 T), $S \approx 0.06$ (7 K, 0.1 T) and $S \approx 0.06$ (15 K, 0.03 T)), and indicate that relaxation by vortex bundles cannot be applied at small fields (above the SF effect). One important difference between our data and the expectation for the SVR is that $J_c(H)$ is not a constant (magnetic field independent) in the range where huge $S$ values are observed. However, a similar field dependence of $J_c$ has been observed in YBCO single crystals with SVR due to random point defects,~\cite{civale94} and attributed to the fast relaxation below the crossover to vortex bundles.~\cite{krusin92} In our case, the experimental $S$ values corresponding to the SVR (above the theoretical prediction for bundles) can be obtained by considering $U_{SVR} $ (2 K) $\approx 30$ K and $\mu = 1/7$.~\cite{blatter94} However, gradual changes in $\mu$ by increasing the magnetic field have been observed in YBCO films with random point defects.~\cite{thompson94} In YBCO, $\mu = 1/7$ is only observed above the SF at magnetic fields where the vortex-vortex interaction is negligible. At higher fields, $\mu$ in the SVR evolves gradually to small vortex bundles.~\cite{civale94}

\begin{figure}[tbp]
\label{figure3}
\centering
\includegraphics[width=8cm]{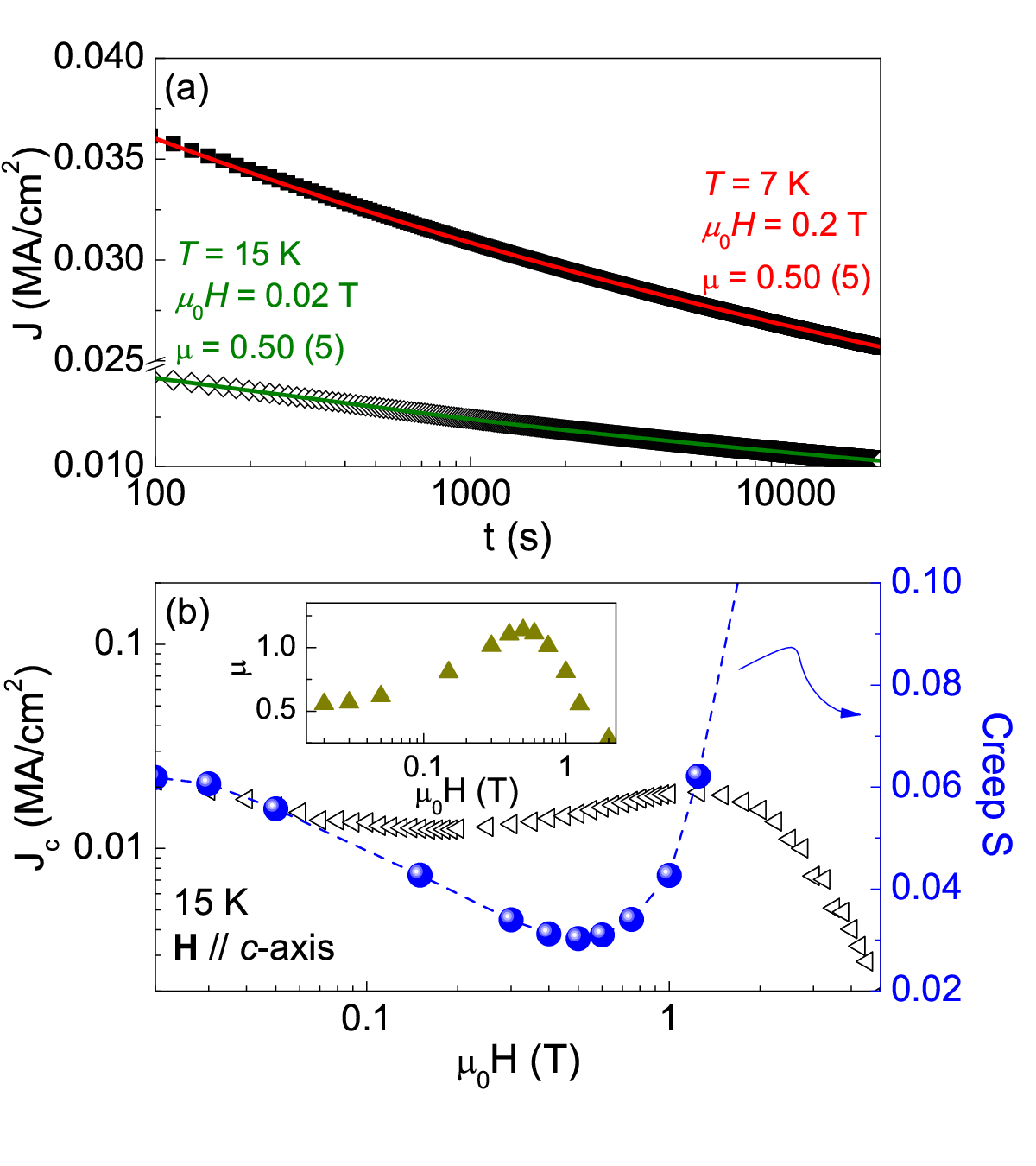}
\caption{(Color online) (a) Persistent current density $J$ vs. ln($t$) at 7 K ($\mu_0H = 0.2$ T) and 15 K ($\mu_0H = 0.02$ T). The red and green lines are fits using the interpolation equation. (b) Magnetic field dependence of $J_c$ and $S$ at 15 K. The inset shows $\mu(H)$ estimation at 15 K.}
\end{figure}

In order to determine the exponent $\mu$, we investigate the nonlogarithmic current decay over a very long period of time ($\approx 16$ hours) at 7 K and 15 K (we assume $T < T_{dp}^S$ and $H > SF$).  Fig. 3(a) shows the time dependence of persistent current density plotted in a semi-logarithmic scale. The curves were adjusted by using the interpolation equation  $J=J_c [\frac {\mu T}{U_c} \mathrm{ln}⁡  \frac {t}{t_0}]^{-1/\mu}$.~\cite{blatter94} The $\mu$ values obtained from the fit in both cases were 0.50 (5). These values are larger than the prediction for the SVR with $\mu = 1/7$ but in the same range as those for YBCO single crystals.~\cite{civale94} Fig. 3(b) shows the $J_c(H)$ and $S(H)$ at 15 K. The inset presents $\mu(H)$ obtained by considering negligible $U_0$ values (see Eq. 1) and bundles $S=1/\mu \mathrm{ln} (t/t_0)$. We select this temperature as it is high enough to avoid self-field problems and low enough to be well within the regime where the second peak in the magnetization exists, so that only a small error in $S$ can be expected. Similarly, the modulation in $J_c(H)$ at 15 K can be understood by considering fast creep relaxation in SVR, and a crossover to vortex bundles. Detailed measurements of $S(H)$ indicate that $\mu$ is smaller ($\approx 1.15$) than the theoretical value $\mu = 3/2$ (see inset Fig 3(b)). Beyond the maximum in $\mu$, a crossover from $sb$ to $lb$ with a gradual reduction to $\mu \approx 7/9$ is observed. Finally the creep changes to a plastic regime above the SPM as was described in ref.[~\cite{prozorov08}]. Nevertheless, our data can be qualitatively described by the collective vortex theory for random point defects, while there is difference in $\mu$. Similar discrepancies, including no discrete $\mu$ values, have been observed in cuprates.~\cite{civale94} It is worth to mention that the theory considers negligible vortex-vortex interaction in comparison with the quenched random potential. In reality, due to the changes in $\lambda(T)$, some interaction between the vortices cannot be completely discarded even at small magnetic fields.~\cite{cho14}

Another scenario to understand the huge $S$ values (higher than the prediction for large vortex bundles) is to consider a non-glassy relaxation by a double kink mechanism.~\cite{niebieskikwiat00} This mechanism appears in YBCO with columnar defects and is manifested as a peak in the relaxation below the matching field $B_{\Phi}$: the density of vortices and defects are the same. The double kink mechanism was ascribed to the presence of different creep regimes: a half vortex loop with $\mu \approx 1$ at low temperatures, superkinks with $\mu \approx 1/3$ in the low-temperature side of the peak up to the maximum in $S(T)$, and a collective regime (bundles) above the peak. The pinning in samples with columnar defects depends on the density and angular distribution of the tracks, the intensity and orientation of the applied field, temperature, and current density. In order to verify the absence of correlated pinning we performed $S$ measurements (not shown) with the magnetic field rotated $10^{\circ}$ away from the $c$-axis. In this configuration high creep rates are expected from the expansion of double kinks. However, no large changes (no increments) in the $S$ values were observed, which rules out the possibility of a double kink mechanism in the sample.~\cite{civale97} This fact is also consistent with the low $J/J_c$ ratio estimated above and with the glassy relaxation observed at different temperatures during the long time measurements.~\cite{niebieskikwiat00}

\begin{figure}[tbp]
\label{figure4}
\centering
\includegraphics[width=8cm]{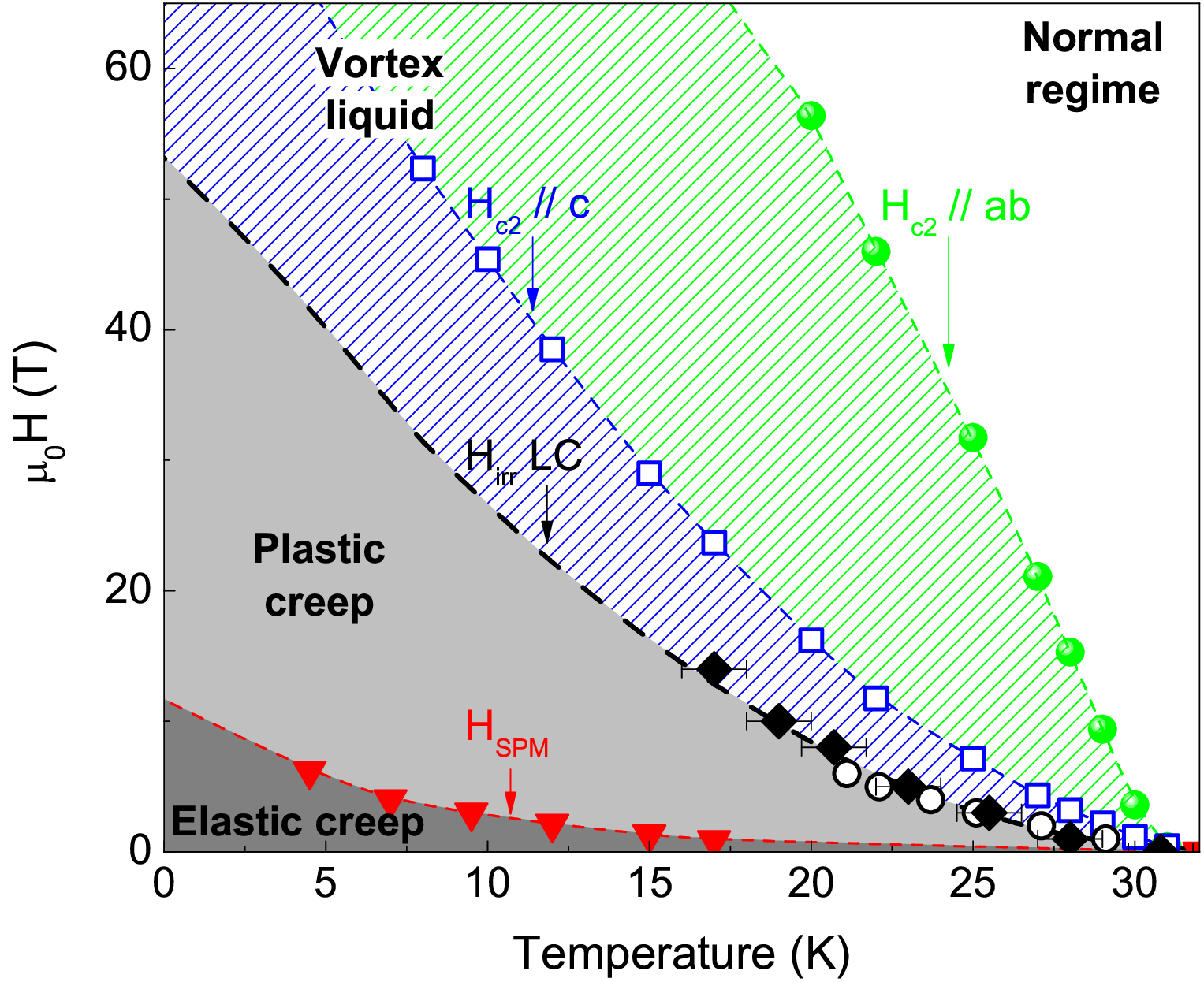}
\caption{(Color online) The $H-T$ phase diagram for superconducting Ca$_{10}$(Pt$_4$As$_8$)[(Fe$_{0.95}$Pt$_{0.05}$)$_2$As$_2$]$_5$ single crystal in the mixed state. Magnetization (open white circles), electrical transport (solid black diamonds), LC: Lindemann criterion, and SPM: second peak in the magnetization.}
\end{figure}

Fig. 4 shows a summary of the $H-T$ vortex phase diagram for the 10-4-8 single crystal. The $H_{c2}(T)$ obtained from PDO for $\mu_0H$ up 60 T shows a similar temperature dependence to that observed in ref.[~\cite{mun12}]. The shapes of the $H_{c2}(T)$ curves for $H//ab$ and $H//c$ close to $T_c$ exhibit the conventional linear field dependence with clearly different slopes for the two field orientations. The $\gamma$ value changes from $\approx 5$ close $T_c$ to $\approx 3$ at 20 K.  A gradual change in $\gamma \rightarrow 1$ is expected by reducing the temperature and increasing the magnetic field above 60 T.~\cite{mun12} At low temperatures, $H_{c2}$(T) with $H//ab$ presents a tendency to saturate, whereas the curve for $H_{c2}$ with $H//c$ shows a steep increase likely associated with two band contribution. Using the same temperature dependence of $H_{c2}(T)$ as in ref.[~\cite{mun12}], $H_{c2}(0)$ values close to 100 T are expected in our sample (see green dotted line $H_{c2}//ab$ in Fig. 4). The irreversibility line ($H_{irr}$) was estimated by considering the irreversibility magnetization and electrical transport (zero resistance).  The $H_{c2}$ and $H_{irr}$ curves spread apart monotonically as $H$ increases. Similar features are observed in high-$T_c$ superconductors where a vortex liquid phase appears.~\cite{blatter94} The $H_{irr}$ line is well described by the Lindemann criterion described above. The fit in Fig. 4 (black dashed line) corresponds to $c_L=0.2$ and $Gi \approx 0.016$, which is in good agreement with the experimental data and suggests a wide vortex liquid phase even at low temperatures. The crossover from elastic to plastic (fast) creep was obtained from the SPM as shown in Fig. 1(a). This line separates elastic motion with a positive $\mu$ to plastic motion with a negative critical exponent.~\cite{prozorov08} A wide region with plastic motion of vortices (fast creep) is identified in the $H-T$ phase diagram. The temperature dependence of the elastic to plastic crossover was adjusted by considering $H^{SPM}(T)=H^{SPM}(0)(1-\frac{T}{T_c})^{n_{exp}}$. The data fits well by considering $H^{SPM}(0) = 10$ T and $n_{exp} = 3.5$. This $n_{exp}$ value is larger than those found for Co-doped BaFe$_2$As$_2$, indicating a sharper temperature dependence of the SPM.~\cite{shen10} Finally, the elastic regime is well described by the collective pinning theory developed for cuprates.~\cite{blatter94} At small fields the vortex dynamics can be described by considering the SVR with a gradual crossover to vortex bundles. It should be noted that the as-grown superconductor with a weak pinning potential is a good candidate to investigate the resulting $H-T$ phase diagram by the inclusion of strong pinning centers.~\cite{civale97}

\section{Conclusion}

We have found that the sample presents a wide vortex liquid phase similar to that predicted in cuprates by considering the Lindemann criteria ($c_L=0.2$ and $Gi \approx 0.016$). The pinning landscape can be associated with random point defects in the whole temperature range ($\sqrt{2} \xi (T) <$ defect radius). The creep $S (H)$ is in agreement with small pinning energies and a gradual crossover from single vortex to relaxation by vortex bundles. Our results are in qualitative agreement with the theory developed for the influence of random point centers on the resulting vortex dynamics of cuprates. In addition, the absolute value of the magnetic penetration depth $\lambda$(4 K) was directly determined in order to understand the role of the intrinsic superconducting properties such as thermal fluctuations in the resulting vortex dynamics.

\begin{acknowledgments}
This work was supported by the Institute for Basic Science (IBS) through the Center for Artificial Low Dimensional Electronic Systems by Project Code (IBS-R015-D1). N. H. is member of CONICET (Argentina). RJ was supported by US NSF DMR-1002622. We thank Jun Sung Kim for a magnetization measurement. We acknowledge the support of the National High Field Laboratory at Los Alamos.
\end{acknowledgments}

\end{document}